\documentclass[12pt,psfig,epsfig,graphicx,psfrag]{article}
\usepackage{dcolumn}
\usepackage{amssymb}
\usepackage{lscape}
\usepackage{amsmath}

\newcommand{\beq}{\begin{equation}}

\newcommand{\eeq}{\end{equation}}
\newcommand{\de}{\partial}

\newcommand{\ud}{\mathrm{d}}
\newcommand{\bea}{\begin{eqnarray}}
\newcommand{\eea}{\end{eqnarray}}
\newcommand{\ba}{\begin{eqnarray}}
\newcommand{\ea}{\end{eqnarray}}

\def\Mf{M_{(5)}}

\def\in{{(i)}}

\def\jn{{(j)}}
\def\kn{{(k)}}

\begin{document}


\begin{flushright}
gr-qc/0505134\\
\end{flushright}
 \vspace{0.5cm}

\begin{center}

 {\Large \bf  Ghosts, Strong Coupling and Accidental Symmetries in \\
\vspace{0.3cm}
 Massive Gravity}\\

 \vspace{0.7cm}

 \vspace{0.5cm}
 {\large
C. Deffayet$^{a,b,}$\footnote{deffayet@iap.fr},
J.-W. Rombouts $^{c,}$\footnote{jwr218@nyu.edu}}\\
\vspace{0.5cm}
$^a${\it APC\;\footnote{UMR 7164 (CNRS, Universit\'e Paris 7, CEA, Observatoire de Paris)}, 11 place Marcelin Berthelot,\\
 75005 Paris Cedex 05, France.}\\
\vspace{0.5cm}
 $^b$ {\it GReCO/IAP\;\footnote{UMR 7095 (CNRS, Universit\'e Paris 6)}
 98 bis boulevard Arago, 75014 Paris, France.}\\
\vspace{0.5cm}
$^c${\it Department of Physics, New York University,\\
 4 Washington Place, New York, NY 10003, USA.}\\
\date{\today}
\vspace{1.5cm}
 \bigskip

{\bf \large Abstract}
\begin{quotation}\noindent
We show that the strong self-interaction of the scalar
polarization of a massive graviton can be understood in terms of
the propagation of an extra ghost-like degree of freedom, thus
relating strong coupling to the sixth degree of freedom discussed
by Boulware and Deser in their Hamiltonian analysis of massive
gravity. This  enables one to understand the Vainshtein recovery
of solutions of massless gravity as being due to the effect of the
exchange of this ghost which gets frozen at distances larger than
the Vainshtein radius. Inside this region, we can trust the
two-field Lagrangian perturbatively, while at larger distances one
can use the higher derivative formulation. We also compare
massive gravity with other models, namely deconstructed theories
of gravity, as well as DGP model. In the latter case we argue that
the Vainshtein recovery process is of different nature, not
involving a ghost degree of freedom.

\end{quotation}

\end{center}


 \newpage

\section{Introduction}
  The construction of a consistent theory of massive gravity has
  proven to be an extremely difficult task. The usual Pauli-Fierz
  approach is to add a quadratic mass term to the quadratic Einstein-Hilbert
  action for a perturbation (the would be massive graviton) over a flat background \cite{Fierz:1939ix}. At this level, the
 action for the massive graviton
  is consistent in the sense that it does not propagate ghosts or
  tachyons. Nevertheless,  it
  is incompatible with experiment because it propagates
  an unwanted extra scalar degree
  of freedom (DOF) that couples to the trace of the energy momentum-tensor
  \cite{vanDam:1970vg} (further referred to as the vDVZ scalar).  It leads
  to physical predictions different from the
  massless theory, even in the limit where the graviton mass
  goes to zero.  This is true for instance for the light bending around
  the sun, and is known as the van
  Dam-Veltman-Zakharov (vDVZ) discontinuity \cite{vanDam:1970vg}.
 An other question raised by the quadratic Pauli-Fierz theory is the link between the background flat metric and the
 graviton. This should be answered by a proper nonlinear theory of massive gravity. However, attempts to go beyond the
 quadratic Pauli-Fierz action are raising even more questions. For example, Boulware and Deser \cite{Boulware:1973my}
introduced an extraneous metric, so that the theory they
considered
 can be thought of as some kind of bimetric theory \cite{Isham:gm}, the dynamics of one of the two metrics
 being frozen, and the coupling between the two metrics being such that it reproduces the quadratic Pauli-Fierz action for
 small perturbations. In the following, we will refer to this type of theory, quite loosely speaking, as massive gravity,
 note however that such a nonlinear completion is not unique. In a particular example of such a theory, simply obtained by
 adding  the quadratic Pauli-Fierz mass term  to
 the full Einstein-Hilbert action, a standard Hamiltonian
 treatment shows that the number of
 propagating DOFs is six (instead of the five DOFs described by
 the quadratic action),
 of which the sixth DOF has a ghost character \cite{Boulware:1973my}. This increment in the number of the propagating DOFs
 with respect to the quadratic theory, as well as the unboundedness from below of the Hamiltonian, was shown
 to
 persist in a more general class of bimetric theories \cite{Damour:2002ws}. This could have consequences at
 the classical level already (see e.g.\cite{Gabadadze:2003jq}).
 On the other hand, it was argued in
  \cite{Deffayet:2001uk,Arkani-Hamed:2002sp} that the vDVZ scalar
  has a strong self interaction on a flat background at an extremely low
  scale $\Lambda=(M_P m^4)^{1/5}$. This, at the classical level
  already,  has dramatic consequences.
  Indeed, Vainshtein
  \cite{Vainshtein:1972sx} first noticed, that nonlinearities in massive gravity
  dominate the linear terms at distance scales smaller than $r_V=(GM/
   m^4)^{1/5}$ (the Vainhstein region) for a classical source of mass $M$.
  He then proposed that these nonlinearities cure the vDVZ discontinuity within
  this region, due to a nonperturbative resummation of the source expansion
  \cite{Vainshtein:1972sx} (see also \cite{Deffayet:2001uk}).
  This "Vainshtein resummation" seems to be problematic for massive gravity
  \cite{Damour:2002gp}, if one follows the resummation beyond the
  first terms, but there remains the possibility it does work
  for more sophisticated models (see e.g. \cite{Deffayet:2001uk}).
  Nonlinearities could also cure some of the above mentioned pathologies by selecting a non asymptotically flat vacuum
  \cite{Damour:2002ws}.

  In this paper, we will only consider expansions of massive gravity over a flat background. Namely we want to discuss the
  relation between
  the strong coupling and the, seemingly unrelated, ghost appearing at nonlinear level in massive gravity. That there
  could   exist a link between these two issues is suggested in part by the following reasoning. Consider e.g. a classical
  non-relativistic source. There it seems, if one follows Vainshtein's reasoning, that
  within the Vainshtein region, one somehow compensates
  for the extra attraction exerted by the vDVZ scalar, that
  is responsible (in a perturbative sense) for the discontinuity (this compensation being at worst only valid far from the
  singularities discussed in \cite{Damour:2002gp}).
  One may suspect that there is a ghost degree of freedom responsible for
  this cancellation\footnote{Such a mechanism was also found in the GRS model
  \cite{Gregory:2000jc}, responsible for the instability of the model
  \cite{Dvali:2000km}.}.
  A natural candidate for this ghost seems to be the "sixth
  degree of freedom" mentioned above, but it is not clear why this ghost
  should only work within a certain distance scale.
  We will discuss in this paper how one can reformulate the strong
  coupling as a ghost problem in the theory, and our analysis
  will clarify the above features.

  Although we will focus here on massive gravity (as
  defined above), we would like to motivate this work by more recent
  developments. Indeed, massive Einstein gravity is just one example
  of a theory that modifies gravity at large distance-scales. In
  recent years, a number of different models have been proposed to
  modify gravity in the infrared. In particular, the
  Dvali-Gabadadze-Porrati (DGP) \cite{Dvali:2000hr} braneworld model (also known as
  {\it brane induced gravity}), modifies gravity at large distance
  while it can give an alternative way of producing the observed late-time cosmic
  acceleration \cite{ced}. DGP gravity shows certain
  similarities with massive gravity. More specifically, it exhibits a
  vDVZ discontinuity at linearized level, and also has a related strong
  coupling \cite{Deffayet:2001uk,Luty:2003vm,Rubakov:2003zb},
  the exact consequences of which has been subject to a debate
  \cite{Deffayet:2001uk,Luty:2003vm,Rubakov:2003zb,Nicolis:2004qq,DEBATE}.
  With the link between strong coupling and ghosts in massive gravity,
  one would hence suspect the DGP model also to contain ghosts
  when properly analyzed (i.e. at the nonlinear level). As we will argue, the situation in DGP is quantitatively
  different from massive gravity, in that the leading operator that
  grows strong in the UV has dimension seven, as opposed to massive
  gravity, where this operator has dimension nine. This fact and the
  particular tensor structure of the operator
  describing the scalar content of the graviton, as found in
  \cite{Luty:2003vm,Nicolis:2004qq}, suggest that the issue is more
  subtle and differs from the case of massive gravity.

  This paper is organized as follows. In the next section we will
  discuss how, in  massive gravity, the strong coupling can be
  reformulated as a ghost problem. Namely we will argue that there is
  a propagating ghost DOF, which appears only at the cubic order in the
   perturbation theory over a flat background, and which is responsible for the
  cancellation, \`a la Vainshtein,  of the attraction exerted by the vDVZ
  scalar around heavy sources.
  We will show how the ghost
  formulation allows a perturbative treatment of massive gravity
  \textit{within} the Vainshtein region (so that in our reformulation, the theory stays
  weakly coupled in this region). This perturbative ghost
  generates the same leading order corrections as found in
  the Vainshtein resummation  as discussed in \cite{Vainshtein:1972sx,Deffayet:2001uk},
  while it freezes out at large distance, where one is only left with the vDVZ scalar.
  This relates the  old Hamiltonian
  formulation to the effective field theoretical formulation. In section 3, we want to
  understand how the appearance of ghosts
  in massive gravity is related
  to the breaking of accidental symmetries, present in the linearized
  theory, at nonlinear level. There we will discuss the link between the ghost discussed in section 2, and the
 "canonical" sixth ghost DOF as discussed by Boulware and Deser.
This section will further give a "geometrical" meaning to the
ghost problem, which we further explore by considering discretized
gravity in section 4. There, we also compare massive gravity with
the DGP model.  We suggest that the situation in the latter is
very
  different as far as the relation between strong coupling and
  ghosts is concerned,
  namely because the operator that grows large \textit{at the lowest energy
  scale} in the DGP model, is not associated with propagating ghost modes.
  We conclude with some final remarks on our work and other IR
  modifications of gravity. Note that our discussion will be
  purely classical, and we will not be concerned with the issue of
  the quantum consistency of the theories considered.

 \section{Ghost or Strong Coupling?} \label{GOSC}
 As recalled above, the quadratic mass term for a spin two excitation is uniquely
 defined by demanding the absence of ghost and tachyonic modes, and takes the Pauli-Fierz
 form \cite{Fierz:1939ix}. Beyond quadratic level, completion is not uniquely defined.
Two possible choices are for example given by the following
actions \bea  S_{BD}&=& M_P^2 \int d^4 x \sqrt{-g} R(g) \nonumber
\\&&+ M_P^2 m^2 \int d^4x \sqrt{-g^{(0)}} h_{\mu \nu} h_ {\alpha
\beta} \left(g_{(0)}^{\mu \nu} g_{(0)}^{\alpha \beta} -
g_{(0)}^{\mu \alpha} g_{(0)}^{\nu \beta}\right), \label{eq1prime} \\
S_{AGS}&=& M_P^2 \int d^4 x \sqrt{-g} R(g) \nonumber \\
&& \label{eq1}+ M_P^2 m^2 \int d^4x \sqrt{-g} h_{\mu \nu} h_
{\alpha \beta} \left(g^{\mu \nu} g^{\alpha \beta} - g^{\mu \alpha}
g^{\nu \beta}\right), \eea
 where
$h_{\mu \nu} \equiv g_{\mu \nu}-g^{(0)}_{\mu \nu}$, $g^{(0)}_{\mu
\nu}$ being some extra metric field\footnote{and $g^{\mu
\nu}_{(0)}$ is the inverse of $g_{\mu \nu}^{(0)}$}. The above
actions define bigravity theories, where the dynamics of
$g^{(0)}_{\mu \nu}$ has been frozen\footnote{This theory is in the
``Pauli-Fierz universality class'', to use the phrasing of ref.
\cite{Damour:2002ws}.}. Note that the above theory is invariant
under 4D diffeomorphisms (in the same sense as bigravity theories
considered in \cite{Isham:gm} are) and arise naturally in the
process of ``deconstructing'' gravity
 \cite{Arkani-Hamed:2002sp,Deffayet:2003zk}.
The action (\ref{eq1prime}) was used by Boulware and Deser in
their Hamiltonian analysis of massive gravity
\cite{Boulware:1973my}, while action (\ref{eq1}) was used by
Arkani-Hamed et al. in the ref. \cite{Arkani-Hamed:2002sp}. In
this last paper, an action for the scalar polarization of the
massive graviton was obtained by introducing the ``hopping''
fields $Y^{\alpha}(x)$, and replacing $h_{\mu \nu}$ by $H_{\mu
\nu}$ given by \beq \label{HOPE}
  H_{\mu \nu}=g_{\mu \nu}-g^{(0)}_{\alpha \beta}(Y) \de_\mu Y^{\alpha} \de_\nu
  Y^{\beta}.\eeq
Taking $g^{(0)}_{\mu \nu}$ and $g_{\mu \nu}$ to
  be flat, and expanding $Y^\alpha$ as $Y^{\alpha}= x^{\alpha}+ \pi^\alpha$, one obtains
 \beq H_{\mu \nu} = h_{\mu \nu} + \pi_{\mu,\nu}
  + \pi_{\nu,\mu} + \pi_{\alpha,\mu}\pi^{\alpha}_{,\nu}.
  \eeq
 Focusing on the scalar mode of the Goldstone vector $\pi^\mu$,
 $ \pi_\mu = \partial_\mu \phi,$ one gets (after an integration by part) the cubic action for $\phi$
\beq
  2 M_P^2 m^2 \int d^4 x \left( (\Box \phi)^3 - (\Box
  \phi)(
  \partial^\mu \partial^\nu \phi)( \partial_\mu \partial_\nu \phi)\right).
 \eeq
Note that, due to general covariance of the Einstein-Hilbert action and to the particular tensorial
structure of the Pauli-Fierz mass term, there is no kinetic term arising directly for $\phi$ from
the above procedure. Rather,   the Goldstone scalar obtains a kinetic term
  only through mixing with  $h_{\mu \nu}$. This gives a $m^2$ dependence (and possibly very small
  coefficient) to the scalar kinetic term $\sim M_{P}^2 m^4 \phi
  \Box \phi$. After a proper diagonalization \cite{Arkani-Hamed:2002sp}, one obtains the following
   action for the canonically normalized scalar (we will still denote $\phi$)
 \ba {\cal L}= \frac{1}{2} \phi \Box \phi +
  \frac{1}{\Lambda^5} (\Box \phi )^3 - \frac{1}{\Lambda^5} (\Box \phi)
  (\partial^\mu \partial^\nu \phi)(\partial_\mu \partial_\nu \phi)
  -\frac{1}{M_P} \phi T, \label{LAGLAG}
 \ea
where $T$ is the trace of the energy-momentum tensor, and the energy scale
  $\Lambda$ is given by
\beq
\Lambda \equiv \left(m^4 M_P \right)^{1/5}.
\eeq
This Lagrangian contains strong cubic interaction terms for the Goldstone
  field, with dimension nine operators growing strong at the scale
  $\Lambda$.
  One can check that the cubic terms are the dominant terms in the effective
  Lagrangian, in the sense that they grow strong at a much
  smaller energy scale than any other interaction terms \cite{Arkani-Hamed:2002sp}, which makes this Lagrangian a useful description of the
  $\phi$ low-energy dynamics. This amounts to taking the limit
  $M_P \to \infty$, $T\to \infty$ and $m \to 0$ with
  $\Lambda= cst$ and $T/M_p =cst$ in the original Lagrangian (\ref{eq1}).
  The limiting procedure eliminates all other self-interaction terms of $\phi$
  as well as mixing terms of the Goldstone scalar with other spin
  components. A similar limit has been considered in
  \cite{Nicolis:2004qq} in the case of DGP gravity. Note also
  that, had we applied the same procedure from action
  (\ref{eq1prime}), we would have obtained the same action as in
  (\ref{LAGLAG}) with a global minus sign in front of the cubic
  derivative interaction terms. This means that the following
  discussion will apply for both cases, and indeed, as we will
  argue, much more generally.

  To investigate the properties of the $\phi$-sector, and with no
  fundamental differences in the conclusion, we simplify our
  discussion here, by omitting the third term in the Lagrangian (\ref{LAGLAG})
  (in the Appendix we explain how to extend the treatment below for
  the full Lagrangian).
  So the starting point in our discussion will be the following
  Lagrangian
  \beq \label{L1} {\cal L}=
 \frac{1}{2} \phi \Box \phi + \frac{1}{ \Lambda^5 } (\Box \phi )^3 -
 \frac{1}{M_P}\phi T.
 \eeq
  This Lagrangian obviously implies an equation of motion of fourth order, reading
 \beq \label{EQMO}
 \Box \phi +\frac{3}{\Lambda^5} \Box( (\Box \phi)^2
 ) - \frac{T}{M_P}=0.
 \eeq
 This means, if one thinks in terms of the Cauchy problem, that one can
  expect this action to
  describe two, rather than one, scalar DOF.
As is well known, similar conclusions are reached in higher
derivative scalar field
  theories (see e.g. \cite{deUrries:1998bi}) or higher
  derivative theories of gravity (see e.g. \cite{Wands:1993uu}).
One can
   typically reformulate
  a higher derivative theory of some fundamental fields as a standard
  two-derivative
  Lagrangian of the fundamental fields plus extra degrees of freedom, with
  certain interaction terms, and which encode the extra derivatives
   appearing in the initial action.
  It is also known that some of these extra DOF, quite generically, have a
  ghost character (see e.g. \cite{Smilga:2004cy}), so that one should be
  able to reinterpret, classically, the strong
  coupling discussed in \cite{Deffayet:2001uk,Arkani-Hamed:2002sp} in terms
  of an extra
  propagating ghost DOF. This is what we do in the following.
  Note that one could question the consequence of this because we consider
  here a truncated
   theory. For example, if one looks at a truncated derivative
   expansion of some perfectly safe underlying theory, one could
   conclude erroneously that the latter is sick. Here, however, the situation
   is quite different because of the fact the Lagrangian with
   started from has some well defined range of applicability, as
   was recalled previously.

  We thus introduce a new field $\lambda$, modifying the Lagrangian
  (\ref{L1}) into
 \ba
 \label{L2}
  {\cal L}_{eq}= \frac{1}{2} \phi \Box \phi + \frac{1}{\Lambda^5}
  (\Box \phi )^3 +F(\lambda,
  \Box \phi)-\frac{1}{M_P}
  \phi T.
 \ea
 The equations of motion for $\lambda$ and $\phi$ are given by
 \beq
  \label{L3}
  \Box \phi +\frac{3}{\Lambda^5} \Box( (\Box \phi)^2 )+\Box F^{(0,1)}
  - \frac{T}{M_P}=0,
 \eeq
 and
 \beq
  \label{L4}
  F^{(1,0)}=0,
 \eeq
 where $F^{(i,j)}$ means the derivative of $F$, $i$ times with respect
 to its first variable and $j$ times with respect to its last one.
 We then ask those equations of motion not to contain derivatives of order
 higher
 than two and to be equivalent to
 equation (\ref{EQMO}). A suitable ansatz for $F$ is given by
 \beq \label{CHOICE}
  F(\lambda, \Box \phi) = \frac{2}{3 \sqrt{3}} \Lambda^{5/2} \lambda^3 + \lambda^2
  \Box \phi - \frac{1}{\Lambda^5} \left(\Box \phi\right)^3,
 \eeq
 leading to the equation of motion
 \ba \label{EQMOEQ1}
  &&\Box \phi + \Box \lambda^2 - \frac{1}{M_P}{T} = 0. \\ \label{EQMOEQ2}
  && \lambda \left(\frac{\Lambda^{5/2}}{\sqrt{3}} \lambda +  \Box \phi \right)= 0.
 \ea
  Those, for a non-vanishing $\lambda$ are equivalent to the equation of motion
  (\ref{EQMO}). Note that the phase space of solutions to the equations
  (\ref{EQMOEQ1}-\ref{EQMOEQ2}) is in fact larger than that of solutions to
  (\ref{EQMO}), because the former also includes solutions to the equation (obtained
  for a vanishing $\lambda$) $\Box \phi + T/M_P =0$. It means that the equivalence we
  are talking about here has to be understood as an equivalence restricted to a suitable
  subset of solutions. Similar restrictions also occur, e.g., when showing the equivalence
  between a $f(R)$ theory of gravity ($f$ being some arbitrary function) and a scalar tensor theory, where the equivalence
   only holds between solutions defined away from the critical points where the second
   derivative of $f$ vanishes \cite{Wands:1993uu} (see also \cite{Hindawi:1995cu}).
  The choice (\ref{CHOICE}) leads to a Lagrangian where the field $\lambda^2$ does get a
   kinetic term through a mixing with $\phi$. Defining then $\varphi$ as
  $\phi = \varphi - \lambda^2$, one obtains the Lagrangian
 \beq
 {\cal L}_{eq}=  \frac{1}{2} \varphi \Box \varphi - \frac{1}{2}\lambda^2 \Box \lambda^2
  + \frac{2}{3\sqrt{3}} \lambda^{3} \Lambda^{5/2} -\frac{1}{M_P} \varphi T+\frac{1}{M_P}
   \lambda^2 T. \label{EQUIVAC1}
 \eeq
 This can be then rewritten as a quadratic action defining $\psi$ as $\psi = \lambda^2$,
 so that ${\cal L}_{eq}$ becomes
 \beq
 {\cal L}_{eq}=  \frac{1}{2} \varphi \Box \varphi - \frac{1}{2}\psi \Box \psi
 -\epsilon \frac{2}{3\sqrt{3}} \psi^{3/2} \Lambda^{5/2} -\frac{1}{M_P} \varphi T+
 \frac{1}{M_P} \psi T, \label{EQUIVAC}
 \eeq
  where $\epsilon = \pm 1 $ is the sign of $\Box(\varphi - \psi)$.
  Notice above the  non-analytic
  form of the potential for $\psi$, as well as the fact that $\psi$ is a positive definite
  field variable. The two different possible choices for $\epsilon$ in the above equation
  indicates that the equivalence between the dynamics obtained for $\phi$ from the Lagrangian
  (\ref{L1}) and that obtained from the Lagrangian (\ref{EQUIVAC}), with a specified value of
  $\epsilon$ is restricted to a set of solutions to the equations of motion having the
  corresponding sign of $\Box \phi \equiv \Box (\varphi - \psi)$. If one consider e.g.
  the Cauchy problem associated with the equation of motion (\ref{EQMO}), and if one
  wants to derive the solution to this Cauchy problem with the equivalent Lagrangian
  (\ref{EQUIVAC}), one should choose the sign $\epsilon$ corresponding to the initial
  Cauchy data provided (of course the sign of $\Box \phi$ can change along the initial
  surface so that the equivalence we are talking about here is really a local property).
  In any case one can also use the Lagrangian (\ref{EQUIVAC1}) where this sign issue does
  not arise, but as we saw, this also leads to solutions of the equations of motion not
  contained in those of (\ref{L1}).

  Thus, we rewrote a Lagrangian of a field with nonrenormalizable interactions
  as a Lagrangian of two degrees of freedom; one free field,
  $\varphi$, corresponding to the
  {\it vDVZ scalar} at large distance from a source, and  one ghost $\psi$,
   with a \textit{relevant} interaction term.

  Let us for now consider a point like source. Then,
  within a certain distance scale to the source, the ghost exactly cancels the
  vDVZ field, up until when the ghost
  freezes out due to its self-interaction, leaving one propagating DOF.
  For our discussion, it is convenient to consider the dimensionless fields
  $\tilde{\varphi} \equiv \varphi/M_P$, and $\tilde{\psi} \equiv \psi/M_P$ which correspond
  with no further normalization factor to the dimensionless massive graviton $h_{\mu \nu}$.
   The equations of motion for those dimensionless fields read
 \ba
 \Box \tilde{\varphi} &=& GT \\
 \Box \tilde{\psi} + \epsilon \frac{m^2}{\sqrt{3}}
 \label{EQMOPSITILDE}
 \tilde{\psi}^{1/2}&=& G T, \ea where $G=\frac{1}{M_P^2}$.
 Considering a classical source with $T=-M \delta^3(x)$, one finds
 that $\varphi$ is given by the usual Newtonian potential (with an
 appropriate sign) \beq \tilde{\varphi}^{(0)} = \frac{GM}{r}, \eeq
 where $r$ is the distance to the source. If one then considers the
 expansion of the solution for $\tilde{\psi}$ around
 $\tilde{\psi}^{(0)} \equiv
  \tilde{\varphi}^{(0)}$, as $\tilde{\psi} = \tilde{\psi}^{(0)} +
  \tilde{\psi}^{(1)} + ... $, one
   finds that $\tilde{\psi}^{(1)}$ obeys to the equation
 \beq \label{PSI1}
  \Box \tilde{\psi}^{(1)}+ \epsilon \frac{m^2}{\sqrt{3}}
  \left(\tilde{\psi}^{(0)}\right)^{1/2} =0,
 \eeq
  which is solved by
 \beq
   \tilde{\psi}^{(1)} = -\epsilon
  \frac{4}{15 \sqrt{3}} m^2 \sqrt{GM} r^{3/2}.
 \eeq
 This matches the correction\footnote{One can verify that
 $\Box \phi$ obtained from this solution has indeed the same sign as
 $\epsilon$. One should choose the sign of $\epsilon$ so that the
 solution has the right asymptotic behavior.} obtained in
 \cite{Vainshtein:1972sx} (see \cite{Deffayet:2001uk}). Here we traced the origin of these
 corrections back to the self-interaction of the perturbative ghost
 $\tilde{\psi}$, which cancels the contribution from the vDVZ scalar
 $\varphi$ at small distances from the source. Notice that
 $\tilde{\psi}^{(0)}$ becomes of the same order as
 $\tilde{\psi}^{(1)}$ at the distance $r_V= (G M/m^4)^{1/5}$, which
 is the Vainstein radius \cite{Vainshtein:1972sx}. Going back to the original field $\phi=\varphi-\psi$, we thus see
 that, inside the Vainshtein region $\phi^{(0)}=0$ and
 \beq
   \phi^{(1)} = \epsilon
   \frac{4}{15 \sqrt{3}} m^2 \sqrt{GM} r^{3/2} \qquad r \ll r_V.
 \eeq
Outside the Vainshtein region, one cannot trust perturbation theory
 for the ghost, but in this region, one can simply use the original
 Lagrangian (\ref{LAGLAG}) for the scalar field perturbatively,
 since the self-interaction of the scalar is small in this region. The
 equations of motion for this Lagrangian read:
 \beq
  \Box \phi +
  \frac{3}{\Lambda^5} \Box (\Box \phi)^2 -
  \frac{1}{\Lambda^5} \Box \left( \partial^\mu
  \partial^\nu \phi
  \partial_\mu \partial_\nu \phi\right) - \frac{2}{\Lambda^5}
  \partial^\mu \partial^\nu \left( (\Box \phi) \partial_\mu
  \partial_\nu \phi \right) - T/M_P = 0
 \eeq
  If we now look for an expansion of the form $\phi = \phi^{(0)} +
  \phi^{(1)}+ ...$ with $\phi^{(1)} \ll \phi^{(0)}$, we see that $\phi^{(0)}$
  obeys
 \beq
  \Box \phi^{(0)} = T /M_P
 \eeq
  so that $\phi^{(0)}$ is given by
 \beq
  \phi^{(0)} = \frac{M}{r M_P}
 \eeq
 for a point like source. $\phi^{(1)}$ is now given by
 \beq
  \Box \phi^{(1)} = \frac{1}{\Lambda^5} \Box \left( \partial^\mu \partial^\nu \phi^{(0)}
 \partial_\mu \partial_\nu \phi^{(0)} \right)
 \eeq
  (the terms in
 $\Box \phi^{(0)}$ vanish outside of the source...) This is solved by
 \beq
  \phi^{(1)} \propto  \frac{M^2}{M_P^3 m^4} \frac{1}{r^6} \qquad
   r \gg r_V,
 \eeq
 which is the form of the correction expected (recall that the
 dimensionless metric fluctuation are given by $\phi/M_P$).

  To summarize, we rewrote the Lagrangian for massive gravity in
  which the vDVZ field gets strongly
  coupled at a scale $\Lambda$, as a system of two
  fields that do not have strong coupling, but
  of whom one is a ghost. In this ghost formulation, we can trust the
  Lagrangian perturbatively within the Vainhstein (or strong
  coupling) region, while outside of the Vainshtein region,
  we can use the original higher derivative interaction
  perturbatively.

\section{Symmetries and Hamiltonian analysis of Massive Gravity}
 The "Goldstone" formalism that we used as a starting point in the previous section is
 especially useful to focus on the sick scalar sector of massive gravity.
 Here we relate our discussion to the original analysis of Boulware and
 Deser,
 who first discussed, in the "unitary gauge",
the unboundedness from below of the Hamiltonian of massive gravity
\cite{Boulware:1973my}. For that discussion, our starting point
will be the action (\ref{eq1prime}), even if, as we already
stressed, the Boulware and Deser findings have been shown to be
valid for generic bimetric gravity theories (and thus massive
gravities as considered here) by Damour and Kogan.

Let us first recall what is going on, from the Hamiltonian point
of view, if one only retains the quadratic part of the action
(\ref{eq1prime}) for the "graviton" $h_{\mu \nu}$. That is to say,
one sets in (\ref{eq1prime}) the background metric $g_{\mu
\nu}^{(0)}$ to be the Minkowski metric, and expands the
Einstein-Hilbert action (first line of (\ref{eq1prime})) to
quadratic order in $h_{\mu \nu}$. One obtains then the Pauli-Fierz
action, which is ghost-free and propagates five degrees of
freedom. The counting of DOFs goes as follows. One start from a
symmetric tensor, the graviton,  which has 10 components. However
it turns out that neither $h_{00}$ nor $h_{0i}$ (with $i=1,2,3$
being spatial indices) are dynamical degrees of freedom. This is a
consequence of the same properties of massless gravity which
shares the same kinetic terms with the theory considered
here\footnote{For the same reason, the $A_0$ component of a
massive Proca fields does not propagates as is the case for the
massless photon}. Namely $h_{00}$ and $h_{0i}$ appear as Lagrange
multipliers in the kinetic term obtained from expanding the
Einstein-Hilbert action. This means that out of the 10 initial DOF
one is left with 10-4= 6 DOF at this stage. Let us then consider
the mass term, here $h_{00}$ and $h_{0i}$ play very different
roles. Indeed the mass term reads  \beq M_P^2 m^2 \label{Massterm}
\int d^4 x \left(h_{ij} h_{ij} - 2 h_{0i}h_{0i} - h_{ii}h_{jj}+ 2
h_{ii}h_{00}\right) \eeq where it appears that $h_{00}$ is a true
Lagrange multiplier for the theory considered, since it enters
linearly both in the kinetic part and mass term of the action. As
a consequence the $h_{00}$ equation of motion generates a
constraint reading \beq \label{CONSHOO}(\nabla^2-m^2) h_{ii}
-h_{ij, ij}=0, \eeq which enables to eliminate one more DOF
leaving 5 propagating DOF in the Pauli-Fierz action. Note that the
equations of motion for the $h_{0i}$, which enter quadratically in
the mass term, do not eliminate other degrees of freedom (in
contrast to the massless case), but rather determines those
variables in terms of the others.

We now turn to fully nonlinear gravity. In the massless case, that
is to say Einstein General Relativity, formulated in the
Hamiltonian language (ADM formalism \cite{Arnowitt:1962hi}), the
lapse and the shift field $N$ and shift $N^{i}$ fields are
Lagrange multipliers associated with the reparametrization
symmetry of the Einstein-Hilbert action \footnote{that is to say,
the Einstein-Hilbert action, in its first order form, is already
in the "parametrized" form, see
e.g.\cite{Arnowitt:1962hi,Hojman:1976vp}}. The  latter are defined
as \beq N\equiv \left(-g^{00}\right)^{-1/2} \eeq and \beq N_i
\equiv g_{0i}, \eeq in terms of the component of the metric
$g_{\mu \nu}$. They generates constraints related respectively to
the time and space reparametrization symmetries of the action.
Those constraints eliminate 4 degrees of freedom, out of the 6
remaining (10-4) leaving the two propagating polarizations of the
massless graviton. The addition of a "mass term" such as the one
of the second line of (\ref{eq1prime}) changes however
dramatically the character of $N$ and $N^{i}$. Indeed, the action
(\ref{eq1prime}) reads in the first order formalism \bea &&
M_P^2\int d^4 x \left\{ \left(\pi^{ij}\dot{g}_{ij}
- N R^0 - N_i R^i\right)\right. \nonumber\\
&& \left.\;\;\;\;\;\;\; - m^2 \left(h_{ij}h_{ij} - 2 N_i N_i -
h_{ii} h_{jj} + 2 h_{ii} \left(1-N^2 + N_k g^{kl} \label{EHADM}
N_l\right)\right) \right\}, \eea where $\pi^{ij}$ are the
conjugate momentum to $g_{ij}$ and $R^0$ and $R^i$ are
respectively the Hamiltonian and momentum constraints of massless
gravity (generated by the lapse and shift fields). The dramatic
observation made in ref. \cite{Boulware:1973my} is that now,
neither $N_i$ nor $N$ are true Lagrange multiplier of the full non
linear massive gravity. Thus, the number if propagating DOF is now
6 rather than five. It is also remarked in ref.
\cite{Boulware:1973my} that the full reduced Hamiltonian is
unbounded from below as we now recall. For this purpose, we
introduce the variable $n$ defined as \beq N\equiv 1+ n \eeq and
rewrites the action (\ref{EHADM}) as \bea && M_P^2\int d^4 x
\left\{ \left(\pi^{ij}\dot{g}_{ij}
- (1+n)R^0 - N_i R^i\right)\right. \nonumber\\
&& \left.\;\;\;\;\;\;\; - m^2 \left(h_{ij}h_{ij} - 2 N_i N_i -
h_{ii} h_{jj} - 2 h_{ii} \left(2n + \beta n^2  - \alpha N_k g^{kl}
\label{EHADM} N_l\right)\right) \right\}, \eea with $\beta$ and
$\alpha$ equal to one. For future reference, we will however keep
explicitly the $\beta$ and $\alpha$ dependence upon the process of
reducing the Hamiltonian. Notice in particular, that upon the
substitution $N_i \rightarrow h_{0i}$ and $2n\rightarrow -h_{00}$,
the mass term (\ref{Massterm}) is obtained from the above
expression by taking $\beta$ and $\alpha$ to vanish. The $N_i$ and
$n$ equation of motion read respectively \bea R^i &=&
4m^2 \left(\eta^{ij} - \alpha h_{kk}g^{ij}\right)N_j \\
R^0 &=& 4 m^2 h_{ii} (1 + \beta n). \eea Those can be used to
extract $n$ and $N^i$ \bea N_j &=& \frac{1}{4 m^2}\left(\eta-
\alpha h_{kk}g\right)^{-1}_{ij} R^i \\ \label{consn}n &=&
\frac{1}{4\beta h_{ii} m^2}\left(R^0 - 4 m^2 h_{jj}\right), \eea
Inserting those expression in action (\ref{EHADM}), we get the
reduced Lagrangian (in first order form), \bea && M_P^2\int d^4 x
\left\{\pi^{ij}\dot{g}_{ij} - m^2 \left(h_{ij}h_{ij} - h_{ii}
h_{jj}\right) - \frac{1}{8m^2} R^l \left(\eta - \alpha h_{ii} g
\right)^{-1}_{lm}R^m \right. \nonumber\\
&& \left.\;\;\;\;\;\;\;\;\;\;\;\;\;\;\;\;- \frac{1}{8 m^2 \beta
h_{ii}}\left(R^0\right)^2 - \frac{2 m^2}{\beta} h_{ii} \right\}.
\label{REDLAG} \eea One reads immediately from the above
expression the Hamiltonian of the system, and discovers that it
can have arbitrary sign and absolute value, since it is true in
particular for the first term of the second line of
(\ref{REDLAG}). This expression (\ref{REDLAG}) shows in a very
clear way that the number of propagating DOF is now six. Note
however that the Hamiltonian is singular when one makes $\beta$ to
vanish. This manifests the fact the sixth DOF only arises from the
integration of $n$, as well as that we have chosen to integrate
the constraint (\ref{consn}) by extracting $n$, which is not
possible if $\beta$ and $h_{ii}$ vanish. So, in this language, the
decoupling of the sixth DOF is subtle and the above Lagrangian can
be considered analogous to the Lagrangian (\ref{EQUIVAC}) where
the two degrees of freedom appear explicitly.

We would like now to relate the unboundedness of the Hamiltonian
for the extra propagating DOF to the analysis done in the section
\ref{GOSC} of this paper. We will discuss this issue in the
covariant formulation, following first the discussion of
\cite{Boulware:1973my}.

Starting from the action (\ref{eq1prime}), one obtains the
following equations of motion for the metric $g_{\mu \nu}$. \beq
\label{CONSCONS} {\cal G}^{\mu \nu} (g=\eta+h) - 2 m^2
\frac{\sqrt{-g_{(0)}}}{\sqrt{-g}} h_{\alpha
\beta}\left(g_{(0)}^{\mu \nu} g_{(0)}^{\alpha \beta} -
g_{(0)}^{\mu \alpha} g_{(0)}^{\nu \beta}\right) = G T^{\mu \nu}
\eeq where ${\cal G}^{\mu \nu}$ is the Einstein tensor for the
metric $g$. In this formulation we can again see how at linearized
level only five DOF appear.
 Indeed, by taking the divergence and trace of the linearized form of the equations of motion,
 one obtains at the linear level
 \beq \label{cstr1}
 \de_{\mu}(h^{\mu \nu} - h \eta^{\mu \nu})=0,
 \eeq
 \beq
 \label{tr}
 h=-\frac{G}{6 m^2} T.
 \eeq
 The first equation implies that the linearized curvature $R^L$ is zero, $R^L$ being given by
\beq R^L = 2\partial_\mu \left(h^{\mu \nu}_{\;\;\; ,\nu} -
\partial^\mu h^\nu_\nu \right). \eeq
The second equation
 states that the trace is not propagating.
The latter constraint, obtained by taking the trace (with respect
to $g_{\mu \nu}$) of equation (\ref{CONSCONS}), generalizes to the
full nonlinear case into
 \beq
 \label{Etr}
 - R-2 m^2
\frac{\sqrt{-g_{(0)}}}{\sqrt{-g}}g_{\mu \nu} h_{\alpha
\beta}\left(g_{(0)}^{\mu \nu} g_{(0)}^{\alpha \beta} -
g_{(0)}^{\mu \alpha} g_{(0)}^{\nu \beta}\right) = G T
\eeq
 which, involving second derivatives, has to be treated as an equation of motion. This is indeed
 where the sixth ghost degree of freedom comes in the covariant formulation.
 Now the first constraint in eq.(\ref{cstr1}) \textit{has} a nonlinear
 generalization which is obtained from the Bianchi Identities for
 the metric $g$. Indeed taking the covariant derivative with
 respect to $g_{\mu \nu}$ of  (\ref{CONSCONS}), one obtains
 \beq
 \left(\frac{\sqrt{-g_{(0)}}}{\sqrt{-g}} h_{\alpha
\beta}\left(g_{(0)}^{\mu \nu} g_{(0)}^{\alpha \beta} -
g_{(0)}^{\mu \alpha} g_{(0)}^{\nu \beta}\right)\right)_{;\mu} = 0.
 \eeq
 Taking the ordinary divergence of this equation allows one to write the linearized curvature
 in terms of higher order terms in $h_{\mu \nu}$. Schematically
 the above equation reads
\beq
 \de_{\nu}\de_{\mu}[h^{\mu \nu}-\eta^{\mu \nu}h]+\Box h^2 +O(h^3)=0,
\eeq
 yielding $ R_L(h)=\Box h^2$ up to terms of $O(h^3)$.
 When plugging back in the trace equation (\ref{Etr}), we get, again schematically, at quadratic order in the field
 equations:
 \beq
  \Box h^2 +m^2 h = G T.
 \eeq
  If we replace $h^2 \to \psi$, we see that this equation has exactly the same form as the equation
  for the ghost of the previous section eq.(\ref{EQMOPSITILDE}).

\section{A comparison with other models}
\subsection{Deconstructed Gravity}
In the previous section, we saw that the Pauli-Fierz mass term was
such that the equivalent of the $R^0=0$ constraint, associated
with time reparametrization of Einstein (massless) General
Relativity, was still present in the massive theory at the level
of the quadratic action and is given by equation (\ref{CONSHOO}).
Thus, somehow, the time reparametrization symmetry is still
present in the quadratic Pauli-Fierz action and eliminates two
degrees of freedom. However it disappears from the fully nonlinear
massive gravity, as we just reminded. It is interesting to remind
a similar property arising in the process of deconstructing
gravity \cite{Arkani-Hamed:2002sp,
Arkani-Hamed:2003vb,Deffayet:2003zk}, where geometry along one (in
the simplest case considered here) dimension is given up by
discretization. The starting point is here the five dimensional
Einstein-Hilbert action reading
 \beq
  \label{ADM} \Mf^3
 \int d^4x dy \sqrt{-g} {\cal N}
 \left\{R + K_{\mu \nu} K_{\alpha \beta}
 \left( g^{\mu \nu} g^{\alpha \beta} - g^{\mu \alpha}
 g^{\nu \beta}\right) \right\},
 \eeq
where we used an ADM-split of the 5D metric along the fifth
space-like dimension to be discretized (see
\cite{Deffayet:2003zk}). One then latticizes the fifth dimension,
with S lattice sites (labelled with latin indices $(i),(j),...$)
with spacing $a$ (we
  will assume periodic identification of the end point of the lattice for
  simplicity). We obtain the discretized action
\bea
\begin{aligned} \label{ORIGINACT}
 S=\Mf^3a\sum_{(i)}\int d^4x\sqrt{-g_{(i)}}{\cal N}_{(i)} & \big[R(g_{(i)})    \\
  +{1 \over 4 {\cal N}_{(i)}^2}
 (\Delta_{\cal L} g_{(i)})_{\mu\nu} & (\Delta_{\cal L} g_{(i)})_{\alpha\beta}
 \big( g^{\mu \nu}_{(i)} g^{\alpha \beta}_{(i)} - g^{\mu
 \alpha}_{(i)}
 g^{\nu \beta}_{(i)}\big)\big],
 \end{aligned}
 \eea
where we took the natural discrete definition of the
Lie-derivative \beq \label{DIFFIN} \Delta_{\cal L} T_{(i)}={
W(i,i+1)T_{(i+1)}-T_{(i)} \over a}. \eeq In this expression $W$ is
the transport operator from site $i+1$ (located at coordinate
$y_{(i+1)}$ along the fifth dimension), to site $i$ (located at
coordinate $y_{(i)}= y_{(i+1)} - a $ along the fifth dimension).
It is given by the Wilson line for transport along the fifth
dimension (in analogy with
  non-abelian gauge theory)
 \beq
  W(y',y)= P \exp \int_y^{y'} dz \bar{N}^\mu\partial_\mu ,\label{def1},
 \eeq
 with $\bar{N}^\mu$ are the
 the shift fields of the ADM-split used in this discretization. The hopping field $Y^{\mu}$
 (analogous to those of equation (\ref{HOPE})) considered in
 \cite{Arkani-Hamed:2002sp} can be explicitly constructed out of W
 as $Y^\mu(y,y_0;x) = W(y,y_0)(x)$.

It turns out to be convenient to work in the Einstein frame for
the metrics on the different sites. Namely, we perform a  Weyl
rescaling $g_{\mu \nu}^\in =
  \exp
  \left(-\frac{\phi_\in}{\sqrt{3}}\right)\gamma_{\mu\nu}^\in$, and
${\cal N}_i \equiv \exp(\phi_\in/\sqrt{3})$, such as the action
(\ref{ORIGINACT}) now reads
\\
 \beq  \label{DISADMbis} \begin{aligned}
 S= \sum_{i} M_p^2
 \int d^4x \sqrt{-\gamma_\in}
 \big\{R(\gamma_\in) -
 \frac{1}{2} \gamma_\in^{\mu \nu} \nabla_{\mu} \phi_\in
 \nabla_{\nu} \phi_\in  &             \\
 + e^{-\sqrt{3} \phi_\in}Q^\in_{\mu \nu} Q^\in_{\alpha \beta}
 \big(\gamma_\in^{\mu \nu} \gamma_\in^{\alpha \beta} &-
 \gamma_\in^{\mu \alpha} \gamma_\in^{\nu \beta} \big) \big\},
 \end{aligned}
 \eeq
 with
 \beq
 \label{DISEXTbis}
 Q^\in_{\mu \nu} = \frac{1}{2} \left\{\Delta_{\cal L} \gamma^\in_{\mu \nu}
   - \gamma^\in_{\mu \nu} \frac{\Delta_{\cal L}
 \phi^\in}{\sqrt{3}}\right\},
 \eeq
 and $M_P^2 = M_{(5)}^3 a$.
Eventually, we want to discuss the symmetries and counting of DOF
of this action at the quadratic level for some fluctuation over a
flat background. We thus expand the different fields as follows
\beq
  \gamma_{\mu\nu}^\in=\eta_{\mu\nu}+{1 \over M_p}
  h_{\mu\nu}^\in,\ \
  \phi^\in={\varphi^\in \over M_p}, \ \
  X^\mu(i,i+1)=x^\mu+{a \over M_p}n^\mu_\in,\label{fluc}
 \eeq
 and define furthermore the discrete Fourier transform of the discrete
  fields as
 \beq
 \hat{\cal F}_\kn=\sum_j{1 \over \sqrt{N}}{\cal F}_\jn
 e^{-i2\pi jk/N}.
 \eeq
 This gives the following quadratic action
 \ba
\begin{aligned}
 S&= \int d^4x \sum_{k} \frac{1}{4}
 \big\{  \partial_\rho \hat h^{\mu
 \nu}_\kn \partial_\sigma \hat h_\kn^{*\alpha \beta}
 \big(   \eta^{\rho
 \sigma}\eta_{\mu \nu} \eta_{\alpha \beta} -\eta^{\rho
 \sigma}\eta_{\mu \alpha} \eta_{\nu \beta} +
 2 \delta^\sigma_{(\nu}
 \eta_{\mu) \beta} \delta^\rho_\alpha
  -  \eta_{\mu \nu} \delta^\sigma_\beta
 \delta^\rho_\alpha \\ &
- \eta_{\alpha \beta} \delta^\sigma_\nu
 \delta^\rho_\mu \big)\big\}
  -{1\over 2}
 \sum_k  \partial_\mu \hat \varphi^\kn \partial_\nu
 \hat \varphi^{*\kn} \eta^{\mu \nu}
 -{1\over 4}
 (\partial_{\mu}\hat n_{\nu}^{(0)}-
 \partial_{\nu}\hat n_{\mu}^{(0)})
 (\partial^{\mu}\hat n^{\nu(0)}-
 \partial^{\nu}\hat n^\mu_{(0)})    \\ &
+\sum_{k\neq 0}
 {1 \over a^2}\sin^2{\pi k\over N} \big\{
 \big.
  \big.
 \big(\big(\hat h_{\mu\nu}^\kn-{\eta_{\mu\nu}\over \sqrt{3}}
 \hat \varphi^\kn\big)-
 {2a\partial_{(\mu} \hat n_{\nu)}^\kn\over e^{i2\pi k/N}-1}
  \big)
 \big( \left(\hat h_{\alpha\beta}^{*\kn}- {\eta_{\alpha\beta}\over
 \sqrt{3}} \hat\varphi^{*\kn}\right)\\  &
 -{2a\partial_{(\alpha} \hat
 n_{\beta)}^{*\kn}\over e^{-i2\pi k/N}-1} \big)
 \left(\eta^{\mu \nu} \eta^{\alpha \beta} - \eta^{\mu \alpha}
 \eta^{\nu \beta}\right) \big\}.
 \end{aligned}
 \label{actionf}
\ea As we will now argue, this action describes, leaving aside
zero modes, a tower of massive spin-two fields with a mass
spectrum given by \beq m^2_k={1 \over a^2}\sin^2{\pi k \over N}.
\eeq Indeed, if one concentrates on the massive modes, the
symmetries of this action are as follows. First one has the
following Stuckelberg symmetry acting at each mass level, namely
\beq \delta\hat h_{\mu\nu}^\kn=
 2\partial_{(\mu} \xi_{\nu)}^\kn,\ \ \delta\hat n_\mu^\kn= {(e^{i2\pi
 k/N}-1)\over a} \xi_{\mu}^\kn\label{invd},\eeq
so that the vector fields
 are the Goldstone bosons which get absorbed by the $(S-1)$ gravitons that
 become massive.
 This symmetry was fully expected from the symmetry of the
 original action (\ref{ORIGINACT}) since it comes from its invariance under the
 product of all the 4D diffeomorphism invariance on each site.
It eliminates $8(S-1)$ DOF out of the $10S + 4(S-1) + S = 15S-4$
DOF present in the action and corresponding respectively to the
$S$ 4D metrics at each sites, the $S-1$ hoping fields $Y$ in
between adjacent sites and the $S$ lapse functions on each site.
However, there is an extra accidental symmetry acting at the
quadratic
 level (in the action), the latter reads
\beq \delta \hat
 h_{\mu\nu}^\kn=\eta_{\mu\nu}f^\kn,\ \delta \hat \varphi^\kn=
 \sqrt{3}f^\kn,\ \delta \hat n_\mu^\kn={a \over 1-e^{-i2\pi k/N}}
 \partial_\mu f ^\kn,\ k\neq 0.\label{invr} \eeq
and eliminates an extra $2(S-1)$
  DOF. We see that this leaves
 $5(S-1)$ propagating DOF (not counting zero modes) corresponding to
 the announced $(S-1)$ massive gravitons.
Interestingly, this symmetry is inherited from the
reparametrization invariance along the discretized direction which
is however broken by the discretization. This is thus analogous to
what was shown to happen for the Pauli-Fierz theory with the time
reparametrization invariance. In particular it is not expected one
can extend this symmetry at the nonlinear level so that extra
degrees of freedom will start propagate there.

\subsection{DGP gravity}
The DGP model is a four dimensionally covariant braneworld model
that is closely related to massive gravity \cite{Dvali:2000hr}.
The DGP setup describes a three brane in a  five dimensional flat
bulk, on which a large Einstein-Hilbert term is present. The
action defining the model is taken to be
\begin{equation}
 S=M_P^2\int{\sqrt{|\bar{g}|} \, \bar{R} \,\ud^4 x} +
M_*^{3} \int{ \sqrt{|g|} \,R \, \ud^{5} X},
\end{equation}
with $M_P \gg M_*$ and where a usual Gibbons-Hawking term taking
care of the brane is implicit. What is important for us is that,
from the point of view of an effective four dimensional brane
observer, the DGP action produces infrared modified Einstein
gravity in a way related to massive gravity. Indeed, in a
particular gauge, the fluctuations around Minkowski vacuum are
described by a Pauli-Fierz alike effective
 action for a massive graviton with "running mass" $m_g^2= p/L_{dgp}$, where
$p$ is the graviton momentum and $L_{dgp}=M_p^2/M_*^{3}$ is the
DGP length scale. This scale was found to be the crossover scale
between a small distance four dimensional behaviour of the
Newtonian potential exchanged by non-relativistic sources on the
brane, and a large distance five dimensional behaviour
\cite{Dvali:2000hr}. As in Pauli-Fierz theory, DGP theory
propagates gravitons with five polarizations at linearized level.
As such, one might expect that the model shares the same
difficulties as massive gravity, especially the strong coupling
and ghost problem. Indeed, the former property of DGP was made
explicit in \cite{Luty:2003vm} and \cite{Rubakov:2003zb} in
confirmation of the work \cite{Deffayet:2001uk} discussing the
appearance of a Vainshtein scale in the model. The exact
consequences of this strong coupling is subject to a debate
\cite{Nicolis:2004qq,DEBATE}. Moreover exact solutions
\cite{Deffayet:2001uk,Kaloper:2005az} as well as approximate ones
\cite{APPROX} indicate that a Vainshtein mechanism is at work in
the DGP model, allowing to recover solutions of Einstein General
Relativity (see however \cite{Gabadadze:2004iy}). In the light of
our previous discussion of massive gravity, one might suspect that
this Vainshtein "resummation" is, also in the DGP model, due to a
dynamical ghost DOF at work. However, with the DGP model being 4D
covariant, one should investigate this issue in detail. There is
one obvious first qualitative difference in between the two
theories. Indeed, the operator in the DGP effective action, which
grows large at a low scale (the strong coupling scale), is a
dimension 7 operator \cite{Luty:2003vm}, as opposed to the
dimension 9 operator found in the the Pauli-Fierz action. The
effective theory for the longitudinal mode of the graviton in DGP
is obtained by decoupling in a similar way to what was recalled in
section \ref{GOSC} for
 massive gravity. It reads  \cite{Nicolis:2004qq}
\begin{equation}
 {\cal L}_{dgp} = 3 \pi \Box \pi -\frac{1}{\Lambda_{dgp}^3}(\de_{\mu} \pi )^2 \Box \pi
 + \frac{ \pi T}{2 M_p}.
\end{equation}
 The scale at which the $\pi$-mode gets strongly coupled is given by $\Lambda_{dgp}=
 (M_p/L_{dgp}^2)^{1/3}$. One might think that the higher
 derivative operator appearing here-above also corresponds to extra degrees of freedom in an equivalent
formulation. However,
 the equation of motion for this scalar are easily seen to be:
\begin{equation}
 6 \Box \pi -\frac{1}{\Lambda_{dgp}^3} (\de_{\mu} \de_{\nu} \pi)^2 +\frac{1}{
 \Lambda_{dgp}^3} (\Box \pi)^2= -\frac{T}{4 M_p}. \label{EQMODGP}
\end{equation}
 Although this is a nonlinear equation of motion, it is a \textit{second order}
 differential equation for the mode
 $\pi$, unlike the analogous equation for its Pauli-Fierz counterpart.
 Hence, as seen from a Cauchy problem perspective, the equation (\ref{EQMODGP}) describes only \textit{one} propagating degree of
 freedom. In \cite{Nicolis:2004qq}, the geometrical meaning of this equation was
 clarified by noting that in the full theory it descends from a combination of
 the Gauss-Codazzi equations and 4D Einstein equations on the brane, which,
 by the relation $K_{\mu \nu} \sim -\frac{1}{\Lambda_{dgp}^3} (\de_{\mu} \de_{\nu}
 \pi)$ makes eqn.(\ref{EQMODGP}) an algebraic equation for the extrinsic curvature of
 the brane. Considering now a point like non-relativistic source,
 as done at the end of section \ref{GOSC}, one finds that inside
 the Vainshtein region, the solution for $\pi$ is given at
 dominant order by the solution to the the quadratic part of
 equation (\ref{EQMODGP}), that is to say the equation obtained by
 dropping the first term in the left hand side of (\ref{EQMODGP})
 \cite{Nicolis:2004qq}. Thus, in order to estimate the contribution
 of $\pi$ to the Newtonian potential around the source, one may say
 that $\pi$ obeyes in this
 region an equation similar to equation (\ref{PSI1}) obeyed by $\tilde{\psi}^{(1)}$, with the r\^ole of
 $\tilde{\psi}^{(0)}$ being played by $T/M_p$. This yields indeed
 corrections to the Schwarzchild solution that goes (correctly) as
 $(r/r^*_v)^{3/2}$, where $r^*_v \equiv \left(r_c^2 GM
 \right)^{1/3}$ is the Vainshtein radius for the DGP model.
 However, here there is no need for the cancellation of an extra
 contribution coming from another DOF.

Note this does not tell us what is the nature of the small
fluctuations around some background solution. In particular, in
\cite{Nicolis:2004qq} it is shown
 that the kinetic term for this scalar is enhanced significantly in the Vainshtein
  region, which incidentally means that the $\pi$-mode interacts more and more
 weakly as one approaches the source.

\section{Conclusion}
In this work, we compared the Goldstone formalism for massive
gravity \cite{Arkani-Hamed:2002sp}, to the old Hamiltonian
approach of Boulware and Deser \cite{Boulware:1973my}.  In the
first approach it was found that the scalar polarization (the vDVZ
scalar) of the graviton acquired a strong cubic interaction, while
in the second approach it was shown that, at nonlinear level,
massive gravity propagates a sixth degree of freedom with negative
energy. We showed that one can reinterpret the strong coupling of
the vDVZ scalar as the propagation of a ghost, in agreement with
the Boulware-Deser finding. The ghost can then be seen as
responsible for the cancellation of the attraction exerted by the
vDVZ scalar at distances around a non-relativistic source smaller
than the Vainshtein's radius. Inside this region, we can trust the
two field Lagrangian perturbatively, while at larger distances one
can use the higher derivative formulation. Of course, within this
region, the presence of the ghost is expected to signal
instabilities. In particular one is tempted to interpret the
failure of the {\it full} Vainshtein's resummation found in
\cite{Damour:2002gp}
 as linked to the presence of this sick DOF, even if one should be
 carefull to draw conclusions from energy arguments in General
 Relativity. Note that the presence of
 this ghost-like DOF is a generic feature of massive
 gravity as shown in the work of Damour and Kogan
 \cite{Damour:2002ws}.

We also compared  massive gravity with other models, namely
deconstructed theories of gravity, as well as the DGP model. In
the latter case we argued that the Vainshtein resummation process
was of different nature, not involving a ghost degree of freedom.
There are other variations on the DGP model which could free of
ghosts too. For instance \cite{Porrati:2004yi} describes a model
in which the five (or higher dimensional) Einstein-Hilbert action
has a profile in the extra dimensions (or alternatively, has a
varying Planck mass in the extra space). In this case, the absence
of ghosts or strong coupling naturally introduces an incurable
vDVZ discontinuity in the theory. We also note that the setup of
"soft massive gravity" \cite{Gabadadze:2003ck} uses higher
dimensional generalizations of the DGP model where the usual
induced Einstein-Hilbert action is supplied with additional UV
operators, and  \cite{Gabadadze:2003ck} argues that it does not
suffer from any strong coupling issues. One interesting other way
to avoid problems of massive gravity might be to break Lorentz
invariance
\cite{Arkani-Hamed:2003uy,Rubakov:2004eb,Dubovsky:2004sg,Gabadadze:2004iv}.

\section*{Acknowledgements}
This work was presented at the PhD defense of J-W.R. on March 21,
2005 at New York University. While this work was being written up,
the preprint hep-th/0505147 appeared on the internet with some
overlap with this work. J-W.R. thanks the GReCO/IAP, where a part
of this research was done, for their warm hospitality. J-W.R. was
supported by a
 Horizon Fellowship of New York University.
We thank G.~Dvali, S.~Dubovsky, G.~Gabadadze, J.~Mourad,
S.~Winitzki for interesting discussions, as well as M.~Porrati for
the same and for encouraging us to publish this paper.

\section*{Appendix: DOF reduction in the general case}
We start from the Lagrangian (\ref{LAGLAG})
\beq \label{L1bis}
  {\cal L}= -\frac{1}{2}
  \de_{\mu} \phi \de^{\mu} \phi + \frac{1}{\Lambda^5} ( (\Box \phi)^3
  - \Box \phi (\de_{\mu} \de_{\nu} \phi)^2 ),
 \eeq
 and want to analyze the degrees of freedom it propagates.
 There are several ways to proceed. There is the "constructive approach" (see e.g.
 \cite{Hindawi:1995cu}, who use similar mechanism to reduce higher
 derivative gravity theories), or the formal
 Ostogradski method for higher derivative scalar field theories, as
 in e.g. \cite{deUrries:1998bi}. Both methods yield in fact the same result,
and we will only discuss here the first one which is in direct
 correspondence with what we did in section \ref{GOSC}.

One can rewrite the Lagrangian (\ref{L1bis}) as
 \beq
 {\cal L}=
 -\frac{1}{2} \de_{\mu} \phi \de^{\mu} \phi + F(\de_{\mu} \de_{\nu}
 \phi), \eeq where $F(X_{\mu \nu})= \frac{1}{\Lambda^5} ( X_{\mu
 \mu}^3 - X_{\mu \mu} X_{\mu \nu}^2 )$.
Next we consider \beq \tilde{{\cal L}}=
 -\frac{1}{2} \de_{\mu} \phi \de^{\mu} \phi + \frac{dF}{dX_{\mu
 \nu}}(\de_{\mu} \de_{\nu} \phi
          -X_{\mu \nu})+ F(X_{\mu \nu}).
 \eeq
 The equations of motion for $X_{\mu \nu}$ and $\phi$ obtained
 from this Lagrangian read
 \bea
0&=&\Box \phi + \partial_\mu \partial_\nu \frac{dF}{dX_{\mu \nu}}
\\
0&=& \frac{d^2F}{dX_{\mu \nu}dX_{\alpha \beta}}\left[\partial_\alpha
\partial_\beta \phi-X_{\alpha \beta}\right].
 \eea
Those equations reduce to those deduced from the Lagrangian
(\ref{L1bis}) if the $16 \times 16$ matrix ${\cal
M}_{(\mu,\nu),(\alpha,\beta)}$ given by \beq {\cal
M}_{(\mu,\nu),(\alpha,\beta)} = \frac{d^2F}{dX_{\mu \nu}
dX_{\kappa \lambda}}, \eeq is invertible. In our case this matrix
is given by \beq \frac{d^2F}{dX_{\mu \nu}dX_{\alpha \beta}} =
\frac{2}{\Lambda^5}\left(3 \eta^{\mu \nu} \eta^{\alpha \beta} X -
\eta^{\mu \nu} X^{\alpha \beta} - \eta^{\alpha \beta} X^{\mu \nu}
- X \eta^{\mu \alpha} \eta^{\nu \beta}\right), \eeq where $X$
stands for $X_{\mu \nu} \eta^{\mu \nu}$. The determinant of the
above defined matrix reads \beq {\rm det}{\cal M} = -2^{18} X^{14}
\Lambda^{-80}\left(4X^2 + X_{\alpha \beta}X^{\alpha \beta}\right),
  \eeq and does not vanish in general. Now we define
 \beq \label{leg}
  \pi^{\mu \nu}=\frac{dF(X)}{dX_{\mu \nu}},
 \eeq
 which we invert to get $X_{\mu \nu}(\pi)$ as a function of
 $\pi^{\mu \nu}$. This inversion is guaranteed to exist again (al least locally) when
  ${\cal M}$ can be inverted, but the explicit form
 of the inverse function may not be easy to obtain. Also one may
 have to divide the theory into different branches if the
function is not globally well defined, in a similar way as what
was done in section \ref{GOSC}. In any case, our new Lagrangian is
now
 \beq
 \tilde{{\cal L}}= -\frac{1}{2} \de_{\mu} \phi \de^{\mu} \phi + \pi^{\mu
 \nu}(\de_{\mu} \de_{\nu} \phi
          -X_{\mu \nu}(\pi))+ G(\pi_{\rho \sigma}),
 \eeq where $ G(\pi_{\rho \sigma}) \equiv F(X_{\mu \nu}(\pi_{\rho
 \sigma}))$.  The function $G$ might now be regarded as a potential
 for the field $\pi_{\rho \sigma}$ (together with $\pi^{\mu
 \nu}X_{\mu \nu}(\pi)$). The investigation of the nature of the
 degree(s) of freedom in $\pi^{\mu \nu}$ in our case simplifies
 drastically because (unlike generic higher derivative gravity for
 example) we have a special mixing with a scalar: the coupling
 $\pi^{\mu \nu} \de_{\mu} \de_{\mu} \phi$ guarantees that there is
 only one component of $\pi^{\mu \nu}$ that propagates due to mixing
 with $\phi$. Indeed, decomposing the symmetric tensor into its
 orthogonal components \beq
  \pi^{\mu \nu}=h_{tt}^{\mu \nu}+\de^{(\mu} A_{t}^{\nu)} + (\eta^{\mu \nu}-
  \frac{\de^{\mu} \de^{\nu}}{\Box}) a +
   \frac{\de^{\mu} \de^{\nu}}{\Box} \lambda,
 \eeq
 where $h_{tt}^{\mu \nu}$ and $A_{t}^{\nu}$ are respectively
 transverse-traceless and transverse, we see that only $\lambda$ gets a kinetic term by mixing with
 $\phi$. The above form of the Lagrangian ensures that $\lambda$ is
 always a ghost. Indeed the Lagrangian is easily diagonalized by the
 transformation $\phi \to \psi=\phi-\lambda$. The final Lagrangian
 for the two DOF is: \beq
  {\cal L}=-\frac{1}{2} \de_{\mu} \psi \de^{\mu} \psi +\frac{1}{2} \de_{\mu}
   \lambda \de^{\mu} \lambda- V(\pi_{\rho \sigma}(\lambda)),
 \eeq
 with $V(\pi_{\rho \sigma})= \pi^{\mu \nu}X_{\mu \nu}(\pi_{\rho
 \sigma})-G(\pi_{\rho \sigma})$. Although the above procedure is a
 bit formal, we did exactly this for the interaction $(\Box \phi)^3$,
 where indeed we encountered two branches and found a potential for
 the ghost $\lambda$ by inverting the scalar analogue of
 Eqn.(\ref{leg}). The exact form of the potential is now harder to
 find, having to solve equation (\ref{leg}) which reads
 \beq
 \label{X}
  \pi^{\mu \nu}=\frac{1}{\Lambda^5}\left(3 \eta^{\mu \nu} X^2 -
  \eta^{\mu \nu} X_{\alpha \beta}X^{\alpha \beta}-2 X X^{\mu
  \nu}\right)
\eeq
 for $X_{\mu \nu}$ as a function of $\pi^{\rho \sigma}$. The least one can say by dimensional analysis
 is that the  potential will be again IR relevant, as opposed to the
 UV relevant operator that we started from in (\ref{L1bis}).

\end{document}